\documentclass[conference]{IEEEtran}
\IEEEoverridecommandlockouts
\usepackage{amsfonts}
\usepackage{setspace}
\usepackage{array}
\usepackage{latexsym}
\usepackage{amssymb}
\usepackage{amsmath}
\usepackage{multirow}
\usepackage[dvips]{graphicx}

\usepackage{subfigure}

\usepackage{url}

\begin{document}

\title{Efficient Quasigroup Block Cipher for Sensor Networks}

\author{\IEEEauthorblockN{Matthew Battey}
\IEEEauthorblockA{Principle Solutions Architect\\
Aspect Software, Inc.\\
Omaha, NE 68164\\
Email: mattbattey@cox.net}
\and
\IEEEauthorblockN{Abhishek Parakh}
\IEEEauthorblockA{Nebraska University Center for Information Assurance\\
University of Nebraska at Omaha\\
Omaha, NE 68182\\
Email: aparakh@unomaha.edu}
}

\maketitle

\begin{abstract}
%\boldmath
We present a new quasigroup based block encryption system with and without cipher-block-chaining. We compare its performance against Advanced Encryption Standard-256 (AES256) bit algorithm using the NIST statistical test suite (NIST-STS) that tests for randomness of a sequence. Since it is well known that a good encryption algorithm must destroy any statistical properties of the input sequence and produce an output close to a true random sequence, the NIST-STS suite results provide a good test bench. In almost all tests from the suite the proposed algorithm performs better than AES256.
\end{abstract}

\section{Introduction}
Sensor networks provide a challenging area of research, because of constraints such as low computational power, low memory capacity and limited communication ranges. As their popularity for various applications such as border surveillance, patient health monitoring, surveillance and environment data collection increases, the demand for security and privacy also increases. Moreover, now a days smart-phones have several sensors within them that may be used to monitor health conditions and used for emergency purposes. In this case, security, integrity and privacy of data that is being transmitted are of utmost importance.

The most popular method for encryption in sensor networks is the use of secret key encryption systems such as Triple DES or AES \cite{parakh1, parakh2}. This is because secret key algorithms have much lower computational requirements compared with public-key systems, which is crucial especially in resource constrained environments such as sensor networks. In this paper we develop a new secret key encryption scheme, that is ideally suited for encryption in low computational and memory constrained environments. We run statistical tests on both the input and output streams, testing them for randomness using the NIST-STS package. The test results are compared with the popularly used Advanced Encryption Standard 256 (AES-256) bit encryption. The results show equal or better performance under all tests and that the encryption method is very good in destroying the structure of the input sequence.

Quasigroups are similar to Sudoku and Latin squares. They have been previously investigated for their application to encryption. Gligoroski et. al. \cite{Gligoroski1, Gligoroski2, Gligoroski3} looked at stream cipher and public key implementations of quasigroups. A multi-level quasigroup implementation was proposed by Satti and Kak \cite{Satti} where they used different sizes of quasigroups to encrypt data. They combined it with indices and nonces to improve on the strength of the encryption. However, their system also focuses on a stream cipher implementation. Marnas et. al. \cite{MarnasBlockQG} implement a quasigroup all-or-nothing system. However, they only use quasigroup encryption to replace the XOR operation used within other all-or-nothing system, hence in the end the actual encryption is done using other cryptosystems. Quasigroups have also been applied to error correction \cite{GligoroskiError} and in construction of message authentication codes (MAC) \cite{BakhtiariMAC}.

One may view quasigroup transformation as a substitution and permutation operation. These transforms form the basis of numerous encryption systems specially in speech encryption \cite{permutation1, permutation2}. Further, public key systems such as NTRU \cite{NTRU} and elliptic curve cryptosystems \cite{ECC} have lower power consumptions compared to RSA however compared to secret key systems they are much more computationally expensive. Moreover, the algorithms proposed in this paper do not require any computations to be performed but only table look up operations for encryption and decryption.

To our knowledge, this is the first quasigroup block encryption algorithm similar in strength to AES with advantages in both computational and memory requirements over the latter.

\section{Background on Quasigroups}
Quasigroups used in cryptography consist of an $n\times n$ matrix consisting of permutations of elements of a finite field $Z_n$ such that no element repeats in any row or column and all elements appear in every row and column. Here $n$ is called the order of the quasigroup. Commonly chosen value of $n$ is 256 such that it allows for us to work with the input stream at byte level. Quasigroups support an operation, denoted by $\cdot$, for any two elements in the matrix such that a corresponding inverse operation, denoted by $\setminus$, exists. For example, for any two elements $x$ and $y$, the following holds true: $x\cdot y = x\cdot z \Rightarrow y=z \mbox{ and } y\cdot x = z\cdot x \Rightarrow y=z$. Further, $x\cdot y = z$ implies $y=x\setminus z$.

The $\cdot$ and $\setminus$ operations are table lookup operations as illustrated by the following example of the working of a conventional quasigroup cipher implementation.

\textbf{Example 1:} Table 1 presents a quasigroup of order 6. The left most column and the top most row are index numbers. An initial seed element is chosen, say $s=3$, and let the input data stream be represented by $\{m_1, m_2, m_3, m_4, m_5, m_6, m_7, m_8\}$ = $\{1,5,4,2,6,4,5,3\}$. Then the encryption process produces an encrypted output stream $\{c_1, c_2, c_3, c_4, c_5, c_6, c_7, c_8\}$ as follows,

\begin{table}
\centering\label{tableQG}
\begin{tabular}{|r|l|l|l|l|l|l|}
\hline
$\cdot$& 1&  2&  3&  4&  5&  6\\ \hline
1& 1&  3&  2&  6&  4&  5\\ \hline
2& 2&  6&  4&  5&  1&  3\\ \hline
3& 3&  2&  6&  4&  5&  1\\ \hline
4& 4&  5&  1&  3&  2&  6\\ \hline
5& 5&  1&  3&  2&  6&  4\\ \hline
6& 6&  4&  5&  1&  3&  2\\ \hline
\end{tabular}
\caption{A quasigroup of order 6.}
\end{table}

\noindent\textbf{Quasigroup Encryption}\\
  \hspace*{1em} 1. Let qGroup[][] represent the quasigroup matrix\\
  \hspace*{1em} 2. To encrypt $m_i$s do,\\
  \hspace*{3em} Set $c_1$ = qGroup[$s$][$m_1$]\\
  \hspace*{3em} For $i>1$, repeat until all $m_i$s are encrypted\\
  \hspace*{4em} $c_i$ = qGroup[$c_{i-1}$][$m_i$]

Execution of the encryption operation for the given input stream is shown below:
\begin{displaymath}
\begin{array}{rcl}
c_1 & = s \cdot m_1  &= 3\cdot 1 = 3 \\
c_2 & = c_1 \cdot m_2 &= 3\cdot 5 = 5 \\
c_3 & = c_2 \cdot m_3 &= 5\cdot 4 = 2 \\
c_4 & = c_3 \cdot m_4 &= 2\cdot 2 = 6 \\
c_5 & = c_4 \cdot m_5 &= 6\cdot 6 = 2 \\
c_6 & = c_5 \cdot m_6 &= 2\cdot 4 = 5 \\
c_7 & = c_6 \cdot m_7 &= 5\cdot 5 = 6 \\
c_8 & = c_7 \cdot m_8 &= 6\cdot 3 = 5 \\
\end{array}
\end{displaymath}

The above encryption operation is a table look up operation over table 1.

For the decryption operation, inverse quasigroup matrix is constructed (table \ref{tab:inverseQG}). To construct the invQGroup[][] matrix, do the following: in the $j^{th}$ column of the $i^{th}$ row in invQGroup[][] matrix write the column number of element $j$ from the $i^{th}$ row in qGroup[][].

\begin{table}
\centering
\begin{tabular}{|r|l|l|l|l|l|l|}
\hline
$\cdot$& 1&  2&  3&  4&  5&  6\\ \hline
1& 1&  3&  2&  5&  6&  4\\ \hline
2& 5&  1&  6&  3&  4&  2\\ \hline
3& 6&  2&  1&  4&  5&  3\\ \hline
4& 3&  5&  4&  1&  2&  6\\ \hline
5& 2&  4&  3&  6&  1&  5\\ \hline
6& 4&  6&  5&  2&  3&  1\\ \hline
\end{tabular}
\caption{Inverse for the quasigroup in Table I.}
\vspace{-0.28in}
\label{tab:inverseQG}
\end{table}

To decrypt do the following,
\begin{enumerate}
  \item $m_1$ = invQGroup[$s$][$c_1$]
  \item For $i>1$, do until all $c_i$s are decrypted
  \begin{itemize}
    \item $m_i$ = invQGroup[$c_{i-1}$][$c_i$]
  \end{itemize}
\end{enumerate}

In general, the direct application of the above encryption algorithm is very effective in randomizing the input data stream. However, given an input data stream and its corresponding output data stream a known plain text attack can be launched because qGroup[$c_{i-1}$][$m_i$] = $c_{i}$. If a long enough mapping is available it may be possible to fill in a significant number of elements in the quasigroup matrix, thereby decreasing the number of possibilities for the group. This is a weakness as the security of the above encryption depends on keeping the quasigroup secret.

\section{Proposed Algorithm 1: Quasigroup Block Cipher}
In order to make quasigroup similar in functionality to the popular AES system, we use 32 different seeds for each round of encryption. Multiple rounds of encryption with different seeds in different rounds finesse the known-plaintext attack and provide a higher level of security, as in the case of Triple DES and AES. We choose 32 seeds, because we assume that each seed is one byte in size and 32 bytes is equal to 256 bits, which is the commonly used key length for AES systems.

In order to introduce dependencies between bytes of input data, we divide the data into 128 bit (16 byte) blocks and encrypt each block separately using Algorithm 1.\\

\noindent\textbf{Algorithm 1}
\begin{enumerate}
  \item Construct a 256x256 size quasigroup.
  \item Generate a random 256 bit encryption key and divide it into 8 bit (1 byte) blocks which will be used as seed elements at every round of encryption. This results in 32, 1 byte, seeds.
  \item Divide the source data into 128 bit (16 byte) blocks
  \item For each block do the following:
  \begin{enumerate}
    \item For each 8-bit block in the cipher key do the following:
    \begin{enumerate}
      \item Using the current block as a stream of 16, 8-bit integers, apply the current 8-bit key as the quasigroup cipher seed and encrypt the block.
      \item Left shift the currently encrypted block by 1, 3, 5 or 7 bits depending on the index of the current 8-bit key block modulo 4.
    \end{enumerate}
  \end{enumerate}
\end{enumerate}

Note that although each block is 128 bits long, when applying quasigroup encryption we further divide the block into 16, 1 byte sub-block. After every round of encryption, all the bits (in the sub-blocks) are taken together and then rotation is applied before the procedure is repeated. A pseudo code is given below:\\

\noindent\verb"Let BlockSize = 16"\\
\verb"Let KeySize = 32"\\
\verb"Define ShiftDistance as [1,3,5,7]"\\
\verb"Define QGMS as Array(256,256)"\\
\verb"Define Key as Array(KeySize)"\\
\verb"Define Source as Array(N, BlockSize)"\\
\verb"Define Output as Array(N, BlockSize)"\\
\verb"For Each Block in Source"\\
\hspace*{1em}\verb"CipherText = Block"\\
\hspace*{1em}\verb"For Each K in Key"\\
    \hspace*{2em}\verb"CipherText = QuasiGroupCipher(QGMS, K,"\\
            \hspace*{4em}\verb"CipherText)"\\
	\hspace*{2em}\verb"CipherText = LeftShift(CipherText,"\\
            \hspace*{4em}\verb"ShiftDistance[IndexOf(K,Key)"\\
            \hspace*{5em}\verb"Modulo 4])"\\
\hspace*{1em}\verb"Next K"\\
\verb"Output[IndexOf(Block,Source)] = CipherText"\\
\verb"Next Block"\\

The shift distances of 1, 3, 5, and 7 are each relatively prime to 2 and thus to 8 (size of a byte). Their sum is 16 (size of 2 bytes) and if each shift is applied 8 times, their sum becomes 128, which is equal to the block size of 128 bits (16 bytes) into which the input data was divided. Therefore, one full rotation of block occurs with shifts of 1, 3, 5 and 7 when all the 32 seeds are used. This ensures that all the bytes in the encrypted block become interdependent.\\

%\begin{figure}[!ht]
%\centering
%\label{fig:flowchart}
%\includegraphics[scale=0.5]{flowchart.eps}
%\caption{Flowchart for the quasigroup block cipher (proposed algorithm 1).}
%\end{figure}

\subsection{Test Implementation}
A test implementation was written in C\#.net, because of the popular adoption of the pre-existing AES cipher suite inbuilt in C\#. Additionally, Microsoft Visual Studio 2010 has built in unit-testing facilities, which combined with Test-Driven-Development, produced well-tested code in reduced increments of time. The test implementation has the ability to overwrite the plaintext buffer, in place, limiting the memory footprint required to encode a buffer. The quasigroup matrix is generated using the Knuth/Fisher-Yates Shuffle \cite{Fisher}. Keys were generated using random-number generator, System.Random, allocating 16 random bytes per request. Both the encryption and decryption routines were constructed and tested.

\subsection{Analysis}
We used the National Institute of Technology - Statistical Test Suite (NIST-STS) suite to evaluate the randomness introduced by the system in the cipher. The NIST-STS package gives a P-value for various standardized tests. The P-value is the probability that a perfect random number generator would have produced a less random sequence than the one being tested \cite{Rukhin}. Control tests were performed against the plain text source. The NIST-STS test suite is available freely in C source code, and downloadable from \url{http://csrc.nist.gov/groups/ST/toolkit/rng/index.html}. The tool can be configured to read a source file as a stream of bits, and evaluate the randomness of that stream. We report the results for the following tests - approximate entropy, block frequency, cumulative sums forward (CS-F) \& cumulative sums reverse (CS-R), fast fourier transform, frequency, longest run, runs, rank and serial 1 and serial 2; where the parameters used for the tests are given in table \ref{NIST-STS_parameters}.

Each test, upon successful completion, produced a P-value result which is to be interpreted as above. If a P-value for a test is determined to be equal to 1, then the sequence appears to have perfect randomness. A P-value of zero indicates that the sequence appears to be completely non-random \cite{Rukhin}. However, both P-values of 1 and 0 are failure conditions in the tests.

\begin{table}\label{NIST-STS_parameters}
\centering
\begin{tabular}{|r|l|l|l|l|l|l|}
\hline
Block Frequency Test - block length(m)& 128\\ \hline
Non-overlapping Template Test - block length(m)& 9\\ \hline
Overlapping Template Test - block length(m)& 9\\ \hline
Approximate Entropy Test - block length(m)& 10\\ \hline
Serial Test - block length(m)& 16\\ \hline
Linear Complexity Test - block length(m)& 500\\ \hline
\end{tabular}
\caption{Parameters for the NIST-STS test}
\vspace{-0.28in}
\end{table}

\begin{table*}
\centering
\begin{tabular}{|p{2cm}|p{1.5cm}|p{1.5cm}|p{2.25cm}||p{1.5cm}|p{1.5cm}|p{1.5cm}|p{1.5cm}|}
\hline
Test & P-value for QG&  P-value for AES&  P-value QG as \% of P-value of AES& All 0x00 input AES& All 0x00 input QG& All 0xFF input AES& All 0xFF input QG\\ \hline
Block Frequency& 0.57189& 0.53593&  106.71& 0.59109& 0.57530& 0.48253& 0.64041\\ \hline
CS-F& 0.47759& 0.45340&  105.33& 0.47739& 0.42955& 0.36766& 0.50679\\ \hline
CS-R& 0.47995& 0.46111&  104.08& 0.48052& 0.43870& 0.36949& 0.49906\\ \hline
FFT& 0.15798& 0.15622&  101.12& 0.03377& 0.043198& 0.05215& 0.05501\\ \hline
Frequency& 0.40314& 0.40006&  100.77& 0.38935& 0.34988& 0.29779& 0.39156\\ \hline
Longest Run& 0.30803& 0.29188&  105.53& 0.24881& 0.21313& 0.17118& 0.27998\\ \hline
Runs& 0.40384& 0.40136& 100.62& 0.37347& 0.37045& 0.38143& 0.35849\\ \hline
\end{tabular}
\caption{The table shows average P-values (over 20 runs) for quasigroup encryption as compared to AES256 encryption system when the same encryption key is used for both cryptosystems without Cipher-Block-Chaining (CBC).}
\vspace{-0.20in}
\label{comparison_aes_qg_NO_CBC}
\end{table*}

Table \ref{comparison_aes_qg_NO_CBC} shows the P-values for the various tests. In the table the first three columns show the average P-values for all zero (0x00) input, all 0xFF input and a text taken from Aesop fables (``From the Goose and the Golden Eggs"). The first column lists the various tests done, second column is the average P-values for encryption of all three inputs using quasigroups, third column is the average P-value for all three inputs using AES and the third column is the ratio of the P-value of encryption using quasigroups to that using AES multiplied by 100. The last four columns are P-values for all zero (0x00) and 0xFF inputs.

\section{Proposed Algorithm 2: Quasigroup Block Encryption with Cipher Block Chaining}

To improve the performance of quasigroup block ciphers in the Approximate Entropy, Serial 1 and Serial 2 tests, we extended algorithm 1 to include cipher block chaining (CBC). Mathematically, CBC is written as:

\[ C_0 := e(k, M_0\oplus iv) \]
\[ C_{n+1} := e(k, M_{n+1}\oplus C_n) \]

Where, $C_n$: an indexed cipher text block, $M_n$: an indexed plain text block, $k$: the cipher key (here seed), $iv$: A random initialization vector, where $|iv|=|C_n|=|M_n|$, $e(k,m)$: the encryption function, QGBC in this case.

\begin{table*}
\centering
\begin{tabular}{|p{2cm}|p{1.5cm}|p{1.5cm}|p{2.25cm}||p{1.5cm}|p{1.5cm}|p{1.5cm}|p{1.5cm}|}
\hline
Test & P-value for QG&  P-value for AES-CBC&  P-value QG as \% of P-value of AES-CBC& All 0x00 input AES-CBC& All 0x00 input QG& All 0xFF input AES-CBC& All 0xFF input QG\\ \hline
Block Frequency& 0.48822&  0.51274&  95.22& 0.52155& 0.47478& 0.50250& 0.48499 \\ \hline
CS-F& 0.51939&  0.50588&  102.67& 0.50527& 0.49851& 0.48968& 0.48843\\ \hline
CS-R& 0.52502&  0.48904&  107.36& 0.49205& 0.51126& 0.47860& 0.49353\\ \hline
FFT& 0.50188&  0.48532&  103.41& 0.46172& 0.48304& 0.49187& 0.49118\\ \hline
Frequency& 0.50190&  0.47353&  105.99& 0.48847& 0.47584& 0.46486& 0.48745\\ \hline
Longest Run& 0.50468&  0.47228&  106.86& 0.47476& 0.46822& 0.46320& 0.53736\\ \hline
Runs& 0.54392&  0.51232&  106.17& 0.53926& 0.55004& 0.51784& 0.54467\\ \hline
Serial 1& 0.53571&  0.53584&  99.98& 0.53300& 0.51054& 0.54146& 0.56533\\ \hline
Serial 2& 0.51635&  0.49246&  104.85& 0.49903& 0.52310& 0.47274& 0.51659\\ \hline
\end{tabular}
\caption{The table shows average P-values (over 20 runs) for quasigroup encryption as compared to AES256 encryption system when the same encryption key is used for both cryptosystems with Cipher-Block-Chaining (CBC).}
\vspace{-0.28in}
\label{table_P-value_with_CBC}
\end{table*}

\subsection{Test Implementation}
After implementing quasigroup block cipher with cipher block chaining, tests were repeated 20 times using a 256 bit random key (32, 1 byte seeds) each time. The resulting encrypted data was tested for randomness using the NIST-STS test suite, using the same parameters as before.

Table \ref{table_P-value_with_CBC} compares a average P-value results from the NIST-STS test suite. The quasigroup block cipher with CBC outperformed AES256 with CBC in almost all cases.

It is to be noted that the variance of P-values between different test results may be misleading, as each test has different acceptance tolerance for P-values.

\subsection{Test on Audio Input}
\begin{figure}[!ht]
\centering
\subfigure{\includegraphics[width=\linewidth]{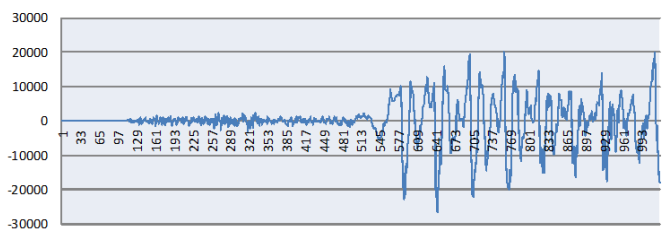}}
\caption{Plot of original input audio waveform}
\label{fig:audioWaves}
\end{figure}

\begin{figure}[!ht]
\centering
\subfigure{\includegraphics[width=\linewidth]{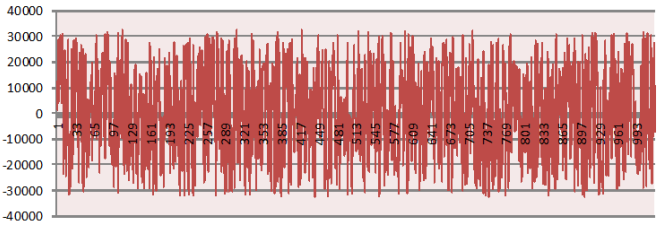}}
\caption{Plot of encrypted output audio waveform}
%\vspace{-0.28in}
\label{fig:audioEnc}
\end{figure}

Since sensors may be used to collect audio signals we perform the encryption operation using quasigroups on an audio input file. The source (taken from \url{http://www.nch.com.au/acm/11k16bitpcm.wav}) and the encrypted audio waveforms are plotted in Figures \ref{fig:audioWaves} and \ref{fig:audioEnc} respectively. As we can see the quasigroup encryption system is very good at distributing the amplitude of the audio signal over the entire range.

We further perform a comparison of the randomness of the signal using the NIST-STS and tabulate the results for the various tests in Table \ref{audio_tests}. We see that in most cases the quasigroup block cipher with CBC randomizes the input waveform much more than AES256 does, especially in the case of Fast Fourier Transform (FFT) tests.

\begin{table}
\centering
\begin{tabular}{|p{2cm}|p{1.5cm}|p{1.5cm}|p{2cm}|}
\hline
Tests& P-value for AES-CBC&  P-value for QG-CBC& P-value of QG-CBC as \% of P-value of AES-CBC\\ \hline
Approximate Entropy& 0.00240& 0.00221& 92.13\\ \hline
Block Frequency& 0.47894&  0.46862& 97.84\\ \hline
CS-F& 0.45362&  0.50949& 112.32\\ \hline
CS-R& 0.45870&  0.49385& 107.66\\ \hline
FFT& 0.44155& 0.49124& 111.26\\ \hline
Frequency& 0.45255&  0.50741&  112.12\\ \hline
Longest Run& 0.47043&  0.48993&  104.15\\ \hline
Rank& 0.48931& 0.46916& 95.88\\ \hline
Runs& 0.48137& 0.49307& 102.43\\ \hline
Serial 1& 0.52025& 0.50224& 96.54\\ \hline
Serial 2& 0.50510& 0.50571& 100.12\\ \hline
\end{tabular}
\caption{P-values for the audio encryption using quasigroup encryption and AES256.}
\label{audio_tests}
\end{table}

\section{On Theoretical Security of Quasigroup Ciphers}
The total number of Latin squares of order $n$, $n>2$, is given by $LS(n)=n!(n-1)!T(n)$, where $T(n)$ denotes the number of reduced Latin squares of order $n$. The numbers $T(n)$ and $LS(n)$ increase very quickly with $n$ \cite{Satti}. Table \ref{latinSquaresTn} gives the number of reduced Latin squares.

\begin{table}
\centering
\begin{tabular}{|r|l|}
\hline
$n$& $T(n)$\\ \hline
2& 1 \\\hline
3& 1 \\\hline
4& 4 \\\hline
5& 56 \\\hline
6& 9048 \\\hline
7& 16942080 \\\hline
8& 535281401585 \\\hline
9& 377597570964258 \\\hline
10& 7580721483160132811489280 \\\hline
11& $5.36\times 10^{33}$\\\hline
12& $1.62\times 10^{44}$\\\hline
13& $2.51\times 10^{56}$\\\hline
14& $2.33\times 10^{70}$\\\hline
15& $1.50\times 10^{86}$\\\hline
\end{tabular}
\caption{Number of reduced latin squares of order 2 to 15.}
\vspace{-0.28in}
\label{latinSquaresTn}
\end{table}

\begin{table}
\centering
\begin{tabular}{|l|}
\hline
$0.689\times 10^{138}\geq LS(16)\geq 0.101\times 10^{119}$\\\hline
$0.985\times 10^{785}\geq LS(32)\geq 0.414\times 10^{726}$\\\hline
$0.176\times 10^{4169}\geq LS(64)\geq 0.133\times 10^{4008}$\\\hline
$0.164\times 10^{21091}\geq LS(128)\geq 0.337\times 10^{20666}$\\\hline
$0.753\times 10^{102805}\geq LS(256)\geq 0.304\times 10^{101724}$\\\hline
\end{tabular}
\caption{Bounds for number of Latin squares for orders 16, 32, 64, 128 and 256.}
%\vspace{-0.28in}
\label{numberOfLatinSquares}
\end{table}

From table \ref{numberOfLatinSquares} we see that the number of possibilities for the Lsatin squares is astronomical. Therefore, if the quasigroup is kept secret along with the 256 bit key (32 random seeds) the system provides very good security.

\section{Conclusion and Future Work}
In this paper we have proposed algorithms for implementation of quasigroup block cipher. The strength of the algorithms was assessed by assessing the randomizing property of the system and the use of statistical test suite by NIST (NIST-STS). Results of the simulations are tabulated and it is observed that in almost all the cases the output generated by the quasigroup encryption system is as or more random than that produced by AES256 for the same encryption key used. The results presented were for average P-values over 20 runs for all zero (0x00) input, all 0xFF input, and an Aesop fable. We also performed tests on audio input and results have been presented.

In future work, we intend to perform cryptanalytic attacks on the proposed quasigroup algorithms. We would also like to make the quasigroup matrix public with only 32 seeds kept secret. A research question would be to see what is a good candidate for a quasigroup (out of numerous possibilities) when it is public \cite{Dvorsky}. We also intend to look into FPGA implementations of the proposed system.

\bibliographystyle{abbrv}
\bibliography{quasigroupICCCNwrkshp_ref}

\end{document}